\documentstyle[preprint,eqsecnum,aps,epsf]{revtex}	

\newif\iftightenlines\tightenlinesfalse
\tightenlines\tightenlinestrue

\begin{document}
%
\def\pT{p_T^{\phantom{7}}}
\def\MW{M_W^{\phantom{7}}}
\def\ET{E_T^{\phantom{7}}}
\def\bh{\bar h}
\def\lm{\,{\rm lm}}
\def\lo{\lambda_1}
\def\lt{\lambda_2}
\def\pslt{p\llap/_T}
\def\eslt{E\llap/_T}
\def\to{\rightarrow}
\def\Re{{\cal R \mskip-4mu \lower.1ex \hbox{\it e}}\,}
\def\Im{{\cal I \mskip-5mu \lower.1ex \hbox{\it m}}\,}
\def\SU{SU(2)$\times$U(1)$_Y$}
\def\te{\tilde e}
\def\tl{\tilde l}
\def\ttau{\tilde \tau}
\def\tg{\tilde g}
\def\tga{\tilde \gamma}
\def\tnu{\tilde\nu}
\def\tell{\tilde\ell}
\def\tq{\tilde q}
\def\tt{\tilde t}
\def\tw{\widetilde W}
\def\tz{\widetilde Z}
\def\cmsec{{\rm cm^{-2}s^{-1}}}
\def\sgn{\mathop{\rm sgn}}
\hyphenation{mssm}
\def\ds{\displaystyle}
\def\ts{${\strut\atop\strut}$}
%
\preprint{\vbox{\baselineskip=14pt%
   \rightline{FSU-HEP-941001}\break
   \rightline{UR-1387}\break
   \rightline{ER-40685-836}
   \rightline{UH-511-807-94}
}}
\title{IMPACT OF R-PARITY VIOLATION ON SUPERSYMMETRY SEARCHES AT THE TEVATRON\\
}
%
\author{Howard Baer$^1$, Chung Kao$^2$ and Xerxes Tata$^3$}
\address{
$^1$Department of Physics,
Florida State University,
Tallahassee, FL 32306, U.S.A.}
\address{
$^2$ Department of Physics and Astronomy, University of Rochester, Rochester,
NY 14627, U.S.A.}

\address{
$^3$Department of Physics and Astronomy,
University of Hawaii,
Honolulu, HI 96822, U.S.A}
\date{\today}
\maketitle
\begin{abstract}
We evaluate cross sections for $\eslt$, 1$\ell$ and various dilepton and
multilepton event topologies that result from the simultaneous production of
all sparticles at the Tevatron collider, both within the minimal model
framework as well as in two different $R$-parity violating scenarios.
Our analysis assumes that these $R$-violating couplings
are small, and that their sole effect is to cause the lightest
supersymmetric particle to decay inside the detector.
We reassess future strategies for sparticle searches at the Tevatron, and
quantify by how much the various signals for supersymmetry could differ from
their minimal model expectations, if $R$-parity is not conserved due to
either baryon number
or lepton number violating operators. We also evaluate the Tevatron
reach in $m_{\tg}$
for the various models, and find that rate-limited
multilepton signals ultimately provide the largest reach for both
$R$-parity conserving and $R$-parity violating cases.

\end{abstract}

\medskip


The search for supersymmetric (SUSY) particles
has become a standard item on the
agenda of experiments at high energy colliders.
Non-observation of events with missing transverse energy ($\eslt$) and
acollinear lepton and/or jet pairs in experiments
at LEP, allows us to infer lower limits\cite{LEP}
$\sim \frac{M_Z}{2}$
on the masses of squarks, sleptons and the charginos. From an analysis of
the $\eslt$ event sample,
the CDF and D0 collaborations at the Tevatron have inferred a
lower limit\cite{TEV} of about 150~GeV on
the masses of gluinos and squarks ($\sim
200~GeV$, if $m_{\tq}=m_{\tg}$). These analyses implicitly
assume that the lightest supersymmetric particle (LSP) is stable and only
weakly interacting, and so
escapes detection in the experimental apparatus yielding the classic $\eslt$
signature for SUSY.
Within the minimal supersymmetric model (MSSM), which is the framework for
most experimental analyses,
the stability of the LSP is guaranteed since there is a multiplicatively
conserved quantum number $R = 1$ (-1) for ordinary particles (sparticles).
It is, however, possible\cite{RPV}
to construct phenomenologically viable models
that do not conserve $R$-parity, but instead conserve either the  baryon
number ($B$) or the lepton number ($L$) (but
not both).
In this case, the LSP decays into ordinary
quarks and leptons, and so, all mass limits based on $\eslt$ analyses
cease to be applicable. The somewhat weaker
bounds\cite{BDT} on sparticle masses from the
measurement\cite{WIDTH} of the $Z$ width at LEP, of course,
continue to be valid.

The phenomenology of $R$-parity violating models can be very different from
that of the MSSM.
$R$-violating interactions, if they are of sufficient
strength, can alter the decay patterns of sparticles from their MSSM
expectations. These interactions also allow sparticles to be produced
singly at colliders, and can lead to resonance production of
squarks or sleptons at the Tevatron\cite{TEVRES,DR} and at HERA\cite{HERA}.
The resulting modifications are sensitively dependent on the strength and
form of $R$-violating interactions, and can be essentially negligible if
these couplings are small relative to the gauge couplings.
Then, the main impact
of $R$-parity violation is, as we mentioned above, that the LSP decays
visibly, invalidating experimental analyses based on the classic $\eslt$
signature. In the clean environment of LEP experiments, it should nonetheless
be possible to search for sparticles by looking for an excess of spherical
events in $Z^0$ decays. In fact, since $Z^0$ decays to LSP pairs can lead
to observable signals if $R$-parity is violated, the non-observation of
spherical events at LEP\cite{OPAL}
translates to a limit $\sim \frac{M_Z}{2}$ on
the mass of the LSP, assuming of course that LSP pair
production is not extremely
suppressed by mixing angle factors. As a result, parameter values
experimentally allowed in LEP experiments may be excluded\cite{GRT}
in an $R$-parity violating scenario.

The corresponding situation at the
Tevatron is quite different. Since the $\eslt$ signals are greatly degraded,
the isolated
multilepton signals from the cascade decays\cite{CAS} of gluinos and squarks
offer
the main hope for the detection of these sparticles at the Tevatron. In the
favourable case where $R$-parity violation is due to $e$ or $\mu$ number
violation, the multilepton signals would be enhanced\cite{DP}. In contrast,
if the LSP decays via $B$-violating interactions, the
additional hadronic activity from LSP decays frequently causes leptons
in SUSY events to fail the lepton isolation criteria, resulting
in a reduction of the multilepton signal. The purpose of this paper is
to quantify how much the various SUSY signals can vary from their canonical
MSSM values if the LSP decays via $R$-parity violating interactions, assuming
that these interactions do not significantly impact either
production rates or decay patterns of sparticles other than the LSP\cite{FN1}.

$R$-parity may be either broken spontaneously (by vacuum expectation
values (VEV) for $R$-odd scalar neutrinos) or explicitly. Spontaneous breaking
via VEVs of the isodoublet sneutrinos of the MSSM is phenomenologically
excluded by measurements\cite{WIDTH} of
$\Gamma_Z$, and so, is only viable if additional
singlet neutrino superfields are introduced. We will, therefore, confine
ourselves to explicit $R$-parity violation via superpotential interactions
which, assuming the MSSM particle content, take the general form,
\begin{eqnarray}
f_{RPV} = \sum_{i,j,k}[\lambda_{ijk}L_iL_jE^c_k + \lambda'_{ijk}L_iQ_jD^c_k
+\lambda''_{ijk}U^c_iD^c_jD^c_k],
\eqnum{1}
\end{eqnarray}
where $i$,$j$ and $k$ denote generations,
and the fields have been defined so that the bilinear lepton number
violating operators have been rotated away. The coupling constants
$\lambda$ ($\lambda''$)
are antisymmetric in the first (last) two indices. The first two terms
lead to lepton number violation, while the last one violates baryon number
conservation. Since the simultaneous presence of both sets of terms would
cause proton decay at a catastrophic rate (unless the couplings are so
tiny as not to be of interest in collider analyses), only $\lambda$ and
$\lambda'$ {\it or} $\lambda''$ type interactions can be present.

The large number of the unknown $R$-parity violating couplings in Eq. (1)
make phenomenological analyses very difficult. In particular,
the decay patterns of the LSP (which, as in the MSSM,
is frequently the lightest neutralino,
$\tz_1$) depend on these couplings. Since we are primarily interested in
exploring the range over which the Tevatron signals vary, we confine our
attention to extreme cases. The only significant published
limit\cite{FN2.5} on $B$-violating $\lambda''$-type
couplings that we are aware of comes from non-observation of $n\bar{n}$
oscillations and requires\cite{ZWIRN}
$\lambda_{112}'',\lambda''_{113} \alt 10^{-6}$.
For the case of $B$-violating interactions, we therefore
assume that the coupling $\lambda''_{212}$, on which there are no significant
experimental constraints\cite{FN3}, dominates LSP decays.
In this case, the LSP decays
via
\begin{eqnarray}
\tz_1\rightarrow cds,\bar{c}\bar{d}\bar{s},
\eqnum{2}
\end{eqnarray}
where CP invariance determines the branching fraction of each of the two
modes to be 50\%. Since we do not attempt to tag $c$ jets, our results
are insensitive to the assumed flavour
structure of this decay. For the case where the LSP decays via lepton number
violating interactions, the multilepton signals are expected to be enhanced.
Since electrons and muons are much easier to detect than tau leptons, we
expect that the enhancement is maximal if the corresponding $L$-violating
interactions involve only $e$ and $\mu$ families. For definiteness, we
assume that the coupling $\lambda_{121}$ dominates, in which case $\tz_1$
decays via,
\begin{eqnarray}
\tz_1\rightarrow \mu\bar{e}\nu_e,
\bar{\mu}e\bar{\nu_e},e\bar{e}\nu_{\mu},e\bar{e}\bar{\nu_{\mu}}.
\eqnum{3}
\end{eqnarray}
Assuming that lepton Yukawa interactions are negligible and that
the sleptons all have the same mass, the four modes each have a branching
fraction of 25\%, independent\cite{GRT} of the gaugino-Higgsino content
of the LSP. We note that $\lambda_{121}$ can be as large as\cite{BGH}
0.08($\frac{m_{\tell}}{200 \ GeV}$) so that the LSP decays well inside the
detector. Constraints on several
other $L$-violating couplings are weaker than those on $\lambda_{121}$.
In these cases, the LSP either decays as in (3) with $\mu$ replaced by
$\tau$ and $\nu_{\mu}$ by $\nu_{\tau}$ (via $\lambda_{131}$ interactions)
or decays via $\tz_1\rightarrow \ell j j,\nu_{\ell}jj$ (via various
$\lambda'$ interactions). The various branching fractions
depend on the parameters of the neutralino mass matrix, and it is not
clear whether these decays will lead to enhancement or degradation of the
signal. What is clear, however, is that any enhancement of the signal will
be smaller than in the case where the LSP decays as in (3), and that the
degradation will be less than that when the LSP decays as in (2). We thus
expect that these cases (Eq. 2 and 3)
represent the extreme limits of SUSY signals at the
Tevatron, assuming only that the $R$-violating interactions are too small
to significantly affect sparticle production mechanisms or the decay patterns
of any sparticles other than the LSP.

The effect of $L$ non-conserving, $R$-parity violating decays of the LSP
on gluino and squark events at the Tevatron was first quantitatively
discussed in
Ref.\cite{DP} using a parton-level Monte Carlo program. It was assumed that
the LSP is a light photino, and further, that gluinos and squarks had only
direct decays to the LSP; {\it i.e.} cascade decays of gluinos and squarks
were ignored. The impact of $L$-violating LSP decays on signals
from $\tw_1\tz_{1,2}$ production at the Tevatron has recently been studied
in Ref.\cite{BBOP}.
Here, we use ISAJET 7.13\cite{ISAJET}
to study the impact of $R$-violating
operators on the MSSM signals, and force the decay of the LSP with branching
fractions discussed above. This improves previous studies in several
respects:
\begin{itemize}
\item ISAJET automatically incorporates the cascade decays of gluinos
and squarks as given by the MSSM.

\item We include contributions from {\it all} SUSY processes that are
kinematically accessible, not just $\tq$ and $\tg$ production. Since
$m_{\tg} >> m_{\tw_1}, m_{\tz_{1,2}}$,
the production of charginos
and neutralinos can make a significant contribution, especially when
gluinos and squarks are heavy.

\item Unlike Ref.\cite{DP} which focussed on the comparison of SUSY
predictions with the Tevatron dilepton data,
we study the impact of $R$-parity violation on all leptonic signals.

\item We also study the impact of baryon number violating operators on the
signal.

\item Finally, ISAJET, which includes effects of radiation
from initial and final states, provides a more realistic simulation
of lepton isolation than a parton-level calculation. This may be especially
important for the discussion of multi-lepton topologies.
\end{itemize}

For our simulation \cite{BKT} of SUSY events, we use CTEQ2L structure
functions\cite{CTEQ}. We model experimental conditions using a toy
calorimeter with segementation $\Delta\eta \times \Delta\phi = 0.1 \times
0.09$ and extending to $|\eta| = 4$. We assume an energy resolution of
$\frac{0.7}{\sqrt{E_T}}$ ($\frac{0.15}{\sqrt{E_T}}$) for the hadronic
(electromagnetic) calorimeter. Jets are defined to be hadron clusters
with $E_T > 15$~GeV in a cone with
$\Delta R=\sqrt{\Delta\eta^2+\Delta\phi^2}=0.7$. Leptons with $p_T > 8$~GeV
and within $|\eta_{\ell}| < 3$ are considered to be isolated if the hadronic
scalar $E_T$ in a cone with $\Delta R = 0.4$ about the lepton is smaller
than $\frac{E_T(\ell)}{4}$.
Finally, since we use the MSSM as the reference model,
we require $\eslt > 20$~GeV in all events. The events are classified as
follows.
\begin{enumerate}
\item For $\eslt$ events, we require $n_{jet} \geq 4$ with at least one of the
jets
in the central region, $|\eta| < 1$, and following the recent analysis by
the D0 collaboration\cite{TEV},
$\eslt \geq 75$~GeV. We veto events
with either isolated leptons with $E_T \geq 15$~GeV (to reduce $W$
backgrounds), or a jet within $30^o$ of $\vec{\eslt}$.
\item Single lepton events are defined to have exactly one isolated
lepton with $E_T
\geq 15$~GeV. We reject events with 60~GeV $\leq m_T(\ell,\eslt) \leq 100$~GeV
which have large backgrounds from $W$ production.
\item The opposite sign (OS) dilepton sample is defined to have two opposite
sign isolated leptons with $p_T \geq 15$~GeV and $30^o \leq
\Delta\phi_{\ell^+\ell'^-} \leq 150^o$ and no other isolated leptons.
To eliminate backgrounds from $Z$
production, we reject events with 80~GeV $\leq m(\ell^+\ell^-) \leq$ 100~GeV.
\item The same sign (SS) dilepton sample is required to have exactly two
isolated leptons, each with $p_T \geq 15$~GeV, and no other isolated leptons.
\item The $n_{\ell} \geq 3$ event sample is defined to have exactly $n_{\ell}$
isolated leptons, with $p_T(\ell_1) \geq 15$~GeV and $p_T(\ell_2) \geq
10$~GeV.
\end{enumerate}

The cross sections for the various SUSY signals calculated within the MSSM
($R$-conserving) framework are shown in Fig.~1 for ({\it a})
$m_{\tq}=m_{\tg}+10$~GeV, ({\it
b}) $m_{\tq}=m_{\tg}-10$~GeV, and ({\it c}) $m_{\tq}=2m_{\tg}$. Here, we
have fixed $\tan\beta=2$, $\mu = -m_{\tg}$ (this is motivated by supergravity
models), $m_t = 170$~GeV, and taken the pseudoscalar Higgs boson mass to be
500~GeV. The slepton masses are determined in terms of $m_{\tg}$
and $m_{\tq}$ using renormalization group equations to evolve from a common
sfermion mass at the GUT scale.
Unlike as in Ref.\cite{BKT}, where
multilepton rates only from $\tg$ and $\tq$
production were shown,
the production of all sparticles at rates expected in the MSSM
is included in this figure. This explains why the ordering of the various
signals sometimes differs from that in Ref.\cite{BKT} and
is also the reason why the curves in Fig. 1
are significantly flatter than those in our previous study --- while the
production of gluinos and squarks dominates for low values of $m_{\tg}$
the production of charginos and neutralinos constitutes $\geq$ 50-90\% of the
total SUSY production cross section (before cuts) if gluinos and squarks are
heavy. Thus, for the very heavy
gluino cases in Fig.~1, we expect that the multilepton signals will be
relatively free of jet activity. This is also the reason why, for large
values of $m_{\tg}$, the rate
for $\eslt$ events (for which we require $n_{jet} \geq 4$) falls below
that of the $1\ell$ event sample on which there is no such
requirement. Finally, we note that the OS and SS dilepton
cross sections in Fig.~1{\it b} increase  sharply for
$m_{\tg}=200-250$~GeV because the decay $\tz_2\rightarrow\tnu\nu$, which is
the only accessible two body decay of $\tz_2$ when $m_{\tg} \alt 200$~GeV,
becomes kinematically forbidden as $m_{\tg}$ is increased from 200~GeV to
250~GeV. As a result, three body leptonic decays of $\tz_2$
(which were negligible for
smaller gluino masses) now add to the dilepton signals, which for
$m_{\tg} \leq 200$~GeV, can come only from chargino decays\cite{BKT}.

The corresponding cross sections in an $R$-parity violating model with
$B$ violation via the $\lambda''_{212}$ coupling ($L$ violation via the
$\lambda_{121}$ coupling) are shown in Fig. 2 (Fig. 3)
for the same three cases of squark mass in Fig. 1. We remind the reader that
these cases should yield the extreme deviations of the SUSY signals
from their MSSM expectations. As before, the
$m_{\tg}=150$~GeV point in Fig.~2{\it b} and Fig.~3{\it b} is excluded
because of constraints on the total width of the $Z$\cite{BDT}, and also,
because these events would lead to novel visible signatures since the LSP is
unstable. We have also checked that the summed branching fraction for Z decays
via $\tz_2\tz_2$,$\tz_1\tz_2$ (and because the $\tz_1$ is visible) or
$\tz_1\tz_1$ is $\sim 4\times 10^{-6}$, which is on the edge of observability
of LEP experiments which have each accumulated a sample of 2M $Z$ events (in
fact, for the lepton number violating case of Fig.~3, this point may well be
already excluded by these data).
The following features of Fig.~2 and Fig.~3 are worthy of note:
\begin{itemize}
\item As expected, the $\eslt$ signal is considerably reduced since, in these
R-violating scenarios, neutrinos are the sole {\it physics} source of $\eslt$.
The
reduction is typically a factor 5-10, but can be as much as two orders of
magnitude in the case of the $L$-violation. In fact, it is interesting to
see that the $\eslt$ event topology has the {\it smallest} cross section
in this L-violating class of models. This is because there are a minimum of
four charged
leptons (from the decays of the two $\tz_1$'s) in each event, and it is rather
unlikely that they all escape detection (recall the lepton veto for the $\eslt$
sample).
\item The ordering of the leptonic signals in $B$-violating models in Fig.~2 is
qualitatively the same as in the MSSM. This is not surprising since the only
difference in the two cases is the hadronic decay of the LSP. Notice that
this additional
hadronic activity makes it harder for the leptons to satisfy the isolation
criteria and results in the anticipated reduction of the leptonic signals.
It is, however, instructive
to note that the signals in Fig.~2{\it a} are generally larger than those
in Fig.~1{\it c}; {\it i.e.} the loss of signal from lepton non-isolation is
smaller than the contribution to the signal from the $\tq\tq$ and $\tq\tg$
production, if $m_{\tq}\sim m_{\tg}$.
\item We see from Fig.~3 that  if the LSP decays exclusively via the
$\lambda_{121}$ coupling in Eq. (1),
the essentially
background-free $3\ell$ and $4\ell$ events will be the dominant SUSY signals
at the Tevatron. Furthermore,
$\sigma(n_{\ell}\geq 3) \geq 0.6 \ pb$ even for $m_{\tg} = 300$~GeV, so
that $\agt 10$ such spectacular events would already be present in the
CDF and D0 data samples, for the set of parameters that we have chosen.
We have not studied the sensitivity of the cross section as a function
of other parameters, and as such, Fig.~3
cannot be taken to mean that
$m_{\tg} \leq 300$~GeV is excluded by experiment. As also pointed out in
Ref.\cite{BBOP} where multilepton
signals from $\tw_1\tz_{1,2}$ production were analysed
within a similar framework, our analysis shows that the
CDF and D0 experiments are indeed probing ranges of parameters not accessible
to them if $R$-parity is conserved. Indeed we see from Fig.~3, that if
sparticle mass patterns are the same as in the MSSM, $\sigma(n_\ell \geq 4)$
exceeds 10 $fb$ for $m_{\tg} \leq 700$~GeV --- we have checked that
the bulk of these events come from $\tw_1\tw_1$ and $\tw_1\tz_2$
production,
which is why the cross section is largest in case ({\it c}) for which
the (negative) interference between the s- and t-channel diagrams is
suppressed\cite{BKTWINO}.
Since events with
$n_{\ell}\geq 4$
are essentially free of SM backgrounds, Tevatron experiments should
be able to indirectly probe gluino masses of 700-800~GeV, after
about one year of Main Injector operation.
It should, however, be kept in mind that
there is no reason for the $\lambda_{121}$ operator to be dominant, so that
the signals may be significantly smaller even in the presence of $L$-violating
LSP decays. Fig.~3 shows just how big these signals can get.
\item We see that unlike as in Fig. 1 and Fig. 2 where $\sigma(\ell^+\ell^-)
> \sigma(\ell^{\pm}\ell^{\pm})$, the OS and SS dilepton
cross sections are essentially equal in Fig. 3. This is because in the
former case $\tz_2$ decays are frequently the dominant source of OS
event topologies, while in the $L$-violating case this signal mainly comes
from the decays of the LSP's, which result in equal amounts of SS and
OS lepton pairs.
For the same reason, OS pairs in Fig.~1 and
Fig.~2 predominantly have the same flavour, whereas in Fig.~3, like and unlike
flavours are equally probable.
\item We should mention that the cuts used in this simulation were motivated
by SUSY searches in the MSSM framework and are not necessarily suitable
for searches when the LSP is unstable. For instance, if the LSP decays via
$\tz_1\rightarrow dcs$,
the requirement $\eslt
\geq 20$~GeV will clearly exclude some
$4\ell$ events where each gluino in a $\tg\tg$ event
decays via $\tg \rightarrow q\bar{q}\tz_2$, $\tz_2\rightarrow
\ell\bar{\ell}\tz_1$. Our purpose here was to study the impact
of $R$-violation on {\it usual} SUSY searches, and not to devise optimal
cuts for $R$-violating scenarios.
\end{itemize}

The {\it physics} backgrounds to these event topologies within the SM
framework are shown in Table I for a top quark mass of 150~GeV and 175~GeV.
We have not attempted to compute detector-dependent backgrounds to multilepton
signals from
misidentification of jets as isolated leptons\cite{KAMON} or to the $\eslt$
signal from mismeasurement of QCD jets which, because of the $\eslt > 75$~GeV
cut, should be small. We see that while SUSY signals and SM backgrounds
are of comparable magnitude in the $\eslt$ and OS dilepton channels,
the signal cross sections substantially exceed backgrounds in the SS and
$n_{\ell} =3$, and in some cases, $n_{\ell} \geq 4$ isolated lepton channels.
We have estimated the reach of the Tevatron by requiring that the SUSY
signal (in any channel) exceed the background by 5$\sigma$; {\it i.e.}
$N_{sig} > 5\sqrt{N_{back}}$, where $N_{sig}$ ($N_{back}$) are
the expected number of signal (background) events in a collider run,
and where we have used the $m_t =150$ GeV background numbers. We
attempt to incorporate systematic uncertainties inherent to these calculations
by further requiring (somewhat arbitrarily) that
$N_{sig}>0.25N_{back}$. We have illustrated the reach of the Tevatron
for the nine cases in Fig.~1--Fig.~3 in Table II,
both for an integrated luminosity
of 0.1 $fb^{-1}$ that is expected to be accumulated by the end of the current
Tevatron run, and, in parenthesis, for an integrated
luminosity of 1 $fb^{-1}$ that
should be accumulated after one year of Main Injector operation. In Table II,
we have required a minimum of five (ten for the Main Injector reach) signal
events in each channel. For the SS and $3\ell$ samples where the expected
background is very small (so that the $5\sigma$ criterion is not meaningful),
we have checked that the Poisson probability for the background to fluctuate
to this minimum event level is $\leq 2\times 10^{-4}$ and $<10^{-5}$,
respectively. Several features of Table II are worth noting:
\begin{itemize}
\item The single lepton signal always appears to be swamped by background
from $W$ production.

\item Even in the MSSM framework, the SUSY reach in the rate-limited
SS and especially the $3\ell$ channels substantially exceeds
the corresponding reach in the $\eslt$ channel (it is possible that the
$\eslt$ reach may
be increased by using a harder $\eslt$ cut\cite{KAMON}) provided a large
enough integrated luminosity can be accumulated, as will be the case at
the Main Injector.

\item For the $B$-violating scenarios in Fig.~2, the $\eslt$ signals are
strongly suppressed (except perhaps in Fig.~2{\it b}); here, the
multilepton signals offers the most promising prospect for SUSY discovery.
It is interesting to see that with the Main Injector, experiments
should be able to
probe values of $m_{\tg}$ up to 200~GeV (350~GeV) if the squarks
are heavy (if $m_{\tq} \sim m_{\tg}$), in the $3\ell$ channel.
Notice that it is possible that
in the worst case scenario of Fig.~2{\it c}, there may be no observable
signal after the current Tevatron run even if the experiments accumulate an
integrated luminosity of 100 $pb^{-1}$ as anticipated. (Values of $m_{\tg}$
substantially below 150~GeV would lead to observable signals from $Z$ decays
at LEP, which is why we do not show the cross sections here.)

\item If instead the LSP decays via the lepton
number violating $\lambda_{121}$ coupling, truly spectacular multilepton
signals would enable experiments at the Tevatron to (indirectly)
probe gluino masses up to $\sim 800$~GeV. We have checked that as much as
about $\frac{1}{3}$ of the $n_{\ell}\geq 4$ events contain 5, or more,
isolated leptons if the gluino is very heavy.
Once again, we emphasize that the magnitude
and event topologies in the $L$-violating case will be sensitively dependent
on the details of the various lepton number violating couplings, and the
results in Fig.~3 should be regarded as upper limits on the ranges of various
signals.
\end{itemize}

In summary, we have examined SUSY signals from the simultaneous production
of all sparticles at the Tevatron. Within the MSSM framework, we find that
at the Main Injector
the multi-lepton signatures should make it possible to probe gluino masses
considerably beyond what can be probed via $\eslt$ searches. We have also
studied
the impact of explicit $R$-parity violating
interactions on
supersymmetry searches at the Tevatron\cite{SUPERC}, assuming that the sole
effect of
these interactions is to cause the LSP to decay inside the detector. If the
$R$-violating couplings are small enough, the LSPs would be rather long-lived,
and their presence in SUSY
events might be inferred by searching for events with displaced vertices.
If this is not the case, the only impact of the $R$-violating LSP decays
would be to alter the cross sections for the various event topologies from
their MSSM values. Most importantly, the cross section for $\eslt$ events is
substantially degraded, so that many experimental lower limits (based on
$\eslt$ searches) are no longer applicable. The large number of independent
$R$-parity
violating couplings that could be present makes a general phenomenological
analysis quite intractable. In this study we have focussed on two models
which, we have argued, cause the maximum variation of the signals from
expectations in the MSSM framework. In the first model, where we assume that
the LSP decays hadronically
via $B$-violating interactions, both $\eslt$ as well as isolated
multi-lepton signals (shown in Fig.~2)
are substantially degraded from their values in the MSSM.
These signals should be relatively insensitive to the assumed
flavour structure of the $B$-violating interactions.
In contrast, SUSY signals are extremely model-dependent if the LSP decays
via lepton number violating interactions.
Multilepton cross sections from SUSY sources
are maximally
enhanced if the LSP decays into a pair of charged leptons ($e$ or $\mu$)
and a neutrino,
as is the case for the model illustrated in Fig.~3. In such a framework,
experiments at the Tevatron, even now, could be probing gluinos as
heavy as 300~GeV, and would be sensitive to gluinos as heavy as
700-800~GeV after the Main Injector upgrade.
It would, however, be really fortuitous
if nature had chosen to
violate $R$-parity in just the right way as to maximize the Tevatron signal;
a more likely situation is to be between the extremes illustrated in
Fig.~2 and Fig.~3. Our projected reach for SUSY searches at the Tevatron
for the various models is summarized in Table II.
The main message of our study is that SUSY may manifest itself quite
differently from MSSM expectations.
There are perfectly viable models where
there may be no observable signal in the $\eslt$ channel but observable
signals in multilepton channels may be present. In fact, even within the
MSSM framework, multilepton channels will provide
the maximum reach in $m_{\tg}$,
once the Main Injector begins operations.
We urge our experimental
colleagues to keep this in mind in designing future SUSY search strategies.

\acknowledgments

We thank H. Dreiner for conversations.
This research was supported in part by the U.~S. Department of Energy
under grant numbers DE-FG-05-87ER40319,
DE-FG-03-94ER40833, and DE-FG-02-91ER40685.
%

%
\newpage

\begin{table}
\caption[]{Standard Model background cross sections in $fb$ for
various event topologies after cuts
described in the text, for $p\bar p$ collisions at $\sqrt{s}=1.8$~TeV.
The $W+jet$ and $Z+jet$ results include decays to $\tau$ leptons.}

\bigskip

\begin{tabular}{ccccccc}
case & $E\llap/_T$ & $1\ \ell$ & $OS$ & $SS$ & $3\ \ell$ & $\ge 4\ \ell$ \\
\tableline
$t\bar t(150)$ & 270 & 1200 & 190 & 0.8 & 0.7 & -- \\
$t\bar t(175)$ & 145 & 590 & 90 & 0.3 & 0.3 & -- \\
$W+jet$ & 710 & $1.2\times 10^6$ & -- & -- & -- & -- \\
$Z+jet$ & 320 & 2200 & 69 & -- & -- & -- \\
$WW$ & 0.4 & 110 & 130 & -- & -- & -- \\
$WZ$ & 0.04 & 4.3 & 1.2 & 2.1 & 0.4 & -- \\
$total\ BG(150)$ & 1300 & $1.2\times 10^6$ & 390 & 2.9 & 1.1 & -- \\
$total\ BG(175)$ & 1175 & $1.2\times 10^6$ & 290 & 2.4 & 0.7 & -- \\
\end{tabular}
\end{table}

\begin{table}
\caption[]{Reach in $m_{\tg}$ via various event topologies
for {\it a}) R-parity conserving (RPC) MSSM,
{\it b}) baryon number violating (BNV) model, and {\it c})
lepton number violating (LNV) model, assuming an integrated
luminosity of 0.1 fb$^{-1}$ (1 fb$^{-1}$), at the Tevatron collider.
We use $m_t =150$ GeV for the background.}

\bigskip

\begin{tabular}{ccccccc}
case & $E\llap/_T$ & $1\ \ell$ & $OS$ & $SS$ & $3\ \ell$ & $\ge 4\ \ell$ \\
\tableline
$a) MSSM$ & & & & & & \\
$m_{\tq}=m_{\tg}+10$ GeV & 240 (260) & --- (---) & 225 (290) &
230 (320) & 290 (425) & 190 (260) \\
$m_{\tq}=m_{\tg}-10$ GeV & 245 (265) & --- (---) & 160 (235) &
180 (325) & 240 (440) & --- (---) \\
$m_{\tq}=2m_{\tg}$ & 185 (200) & --- (---) & --- (180) &
160 (210) & 180 (260) & --- (---) \\
\tableline
$b) BNV$ & & & & & & \\
$m_{\tq}=m_{\tg}+10$ GeV & --- (---) & --- (---) & 165 (210) &
200 (280) & 220 (350) & --- (165) \\
$m_{\tq}=m_{\tg}-10$ GeV & 200 (210) & --- (---) & 150 (165) &
165 (235) & --- (360) & --- (---) \\
$m_{\tq}=2m_{\tg}$ & --- (---) & --- (---) & --- (---) &
--- (200) & --- (190) & --- (---) \\
\tableline
$c) LNV$ & & & & & & \\
$m_{\tq}=m_{\tg}+10$ GeV & --- (150) & --- (---) & 240 (300) &
330 (450) & 480 (650) & 540 (740) \\
$m_{\tq}=m_{\tg}-10$ GeV & 160 (180) & --- (---) & 250 (300) &
330 (450) & 460 (640) & 520 (710) \\
$m_{\tq}=2m_{\tg}$ & --- (---) & --- (---) & 190 (260) &
340 (540) & 540 (730) & 600 (840) \\
\end{tabular}
\end{table}
%
\newpage
\begin{figure}
\caption[]{Cross sections at the Tevatron ($\sqrt{s}=1.8$ TeV)
in {\it fb} for various event topologies after
cuts given in the text for the $R$-parity conserving MSSM,
for three choices of
squark mass. We take $|\mu |=-m_{\tg}$, $\tan\beta =2$, $A_t=A_b=-m_{\tq}$
and $m_{H_p}=500$ GeV. The $\eslt$ events are labelled with diamonds, the
1-$\ell$ events with crosses, the $\ell^+\ell^-$ events with x's and
the $SS$ with squares. The dotted curves are for 3-$\ell$ signals, while
dashes label the 4-$\ell$ signals. For clarity, error bars are
shown only on the lowest lying curve; on the other curves the error bars are
considerably smaller. We note that the $m_{\tg}= 150$~GeV case in Fig. 1{\it b}
is already excluded by LEP constraints on the $Z$ width, since this implies
$m_{\tnu}=26$~GeV.}
\end{figure}

\begin{figure}
\caption[]{Same as Fig. 1, except that the LSP is assumed to decay via
$\tz_1\to cds$ or $\bar c\bar d\bar s$ due to the $R$-parity violating
$\lambda''_{212}$ coupling.}
\end{figure}

\begin{figure}
\caption[]{Same as Fig. 1, except here $\tz_1\to \mu\bar e\nu_e$,
$\bar\mu e\bar\nu_e$, $e\bar e\nu_{\mu}$ or $e\bar e\bar{\nu_{\mu}}$,
each 25\% and that the dashed curve includes $n_{\ell}\geq 4$ events.
The multilepton cross sections shown here should be interpreted as
upper limits on cross sections in models where the sole effect of $R$-parity
violation is to cause the LSP to decay inside the detector. The
$m_{\tg}=150$~GeV points may already be excluded by LEP data as discussed in
the text.}
\end{figure}


\begin{references}
\bibitem{LEP} D.~Decamp {\it et.al.} (ALEPH Collaboration),
Phys.~Rep. {\bf 216}, 253 (1992);
P.~Abreu {\it et.al.} (DELPHI Collaboration),
Phys.~Lett. {\bf B247}, 157 (1990);
O.~Adriani {\it et.al.} (L3 Collaboration),
Phys.~Rep. {\bf 236},1 (1993);
M.~Akrawy {\it et.al.} (OPAL Collaboration),
Phys. Lett {\bf B240}, 261 (1990); for a review, see
G. Giacomelli and P. Giacomelli, Riv. Nuovo Cim. {\bf 16}, 1 (1993).
\bibitem{TEV} M.~Paterno, Ph.D. thesis; D.~Claes (D0 Collaboration),
{\it presented at the 8th
DPF meeting, Albuquerque, NM, Aug. 1994}; F.~Abe {\it et. al.},
(CDF Collaboration), Phys. Rev. Lett. {\bf 69}, 3439 (1992).
\bibitem{RPV} C.~S.~Aulakh and R.~N.~Mohapatra, Phys. Lett. {\bf B119}, 316
(1982); L.~J.~Hall and M.~Suzuki, Nucl. Phys. {\bf B231}, 419 (1984);
S.~Dawson, Nucl. Phys. {\bf B261}, 297 (1985); S. Dimopoulos and L.~Hall,
Phys. Lett. {\bf B207}, 210 (1987); L.~Hall, Mod. Phys. Lett. {\bf A5}, 467
(1990).
\bibitem{BDT}
See {\it e.g.} H.~Baer, M.~Drees and X.~Tata,
Phys. Rev. {\bf D41}, 3414 (1991).
\bibitem{WIDTH} S. Olsen, {\it Plenary talk at the 8th
DPF meeting, Albuquerque, NM, Aug. 1994}.
\bibitem{TEVRES} S.~Dimopoulos {\it et. al.} Phys. Rev. {\bf D41}, 2099
(1990).
\bibitem{DR} H.~Dreiner and G.~Ross, Nucl. Phys. {\bf B365}, 597 (1991).
\bibitem{HERA} J.~Butterworth and H.~Dreiner, Nucl. Phys. {\bf B397}, 3
(1993).
\bibitem{OPAL} Bounds on unstable photinos which can be produced
at LEP via $t$-channel
selectron exchange have been
studied in the context of an $L$ violating model by
P. Acton {\it et. al.} (OPAL Collaboration), Phys. Lett.
{\bf B313}, 333 (1993).
\bibitem{GRT} R.~Godbole, P.~Roy and X.~Tata, Nucl. Phys. {\bf B401}, 67
(1992).
\bibitem{CAS} H.~Baer {\it et. al.}, Phys. Lett. {\bf B161},175 (1985);
G.~Gamberini. Z. Phys. {\bf C30}, 605 (1986); H. Baer, V. Barger, D. Karatas
and X. Tata, Phys. Rev. {\bf D36}, 96 (1987); G. Gamberini, G. Giudice,
B. Mele and G. Ridolfi, Phys. Lett. {\bf 203B}, 453 (1988);
R. M. Barnett, J. Gunion and H. Haber, Phys. Rev. {\bf D37}, 1892 (1988);
A. Bartl, W. Majerotto, B. Mosslacher and N. Oshimo,
Z. Phys. {\bf C52}, 477 (1991).
\bibitem{DP} D.~P.~Roy, Phys. Lett. {\bf B283}, 270 (1992).
\bibitem{FN1} If all $R$-parity violating couplings are small enough, they
do not make significant contributions (via renormalization effects) to
sparticle masses and mixings. Thus sparticle mass patterns are not
significantly modified from minimal model expectations.
\bibitem{FN2.5} It was thought that
cosmological arguments requiring GUT-scale baryogenesis not
get washed out require that the $\lambda''$-type couplings are $\alt 10^{-7}$.
It has since been shown (see {\it e.g.} H.~Dreiner and G.~Ross, Nucl. Phys.
{\bf B210}, 188 (1993)) that these bounds are model-dependent and can be
evaded in reasonable scenarios.
\bibitem{ZWIRN} F.~Zwirner, Phys. Lett. {\bf B132}. 103 (1983); R.~Barbieri
and A.~Masiero, Nucl. Phys. {\bf B267}, 679 (1986).
\bibitem{FN3} Because of intergenerational mixing there should
be (weaker) bounds on
the other $\lambda''$ couplings (especially on
$\lambda''_{212}$) than the one on
$\lambda''_{112}$ obtained in Ref.\cite{ZWIRN}.
The important thing is that several of these couplings may be large enough
so that the LSP would decay within
the detector. There is essentially no experimental bound
on the $\lambda''_{223}$ coupling (upper bounds obtained by requiring
that the $B$-violating couplings remain perturbative up to a GUT
scale do not lead to interesting phenomenological
constraints, as discussed by B.~Brahmachari
and P.~Roy, Phys. Rev. {\bf D50}, 39 (1994)).
If this coupling dominates $B$-violating
interactions, signatures would be very similar to the ones we discuss, with
the $d$-quark in the LSP being replaced by a $b$. Except that an accidently
non-isolated lepton would occasionally contribute to the leptonic signals
discussed below, the signals we discuss are quite insensitive to
the flavour structure of the $B$-violating interaction (we do not consider
the possibility of tagging the $b$-quarks from LSP decay).
\bibitem{BGH} V.~Barger, G.~Giudice and T.~Han, Phys. Rev. {\bf D40}, 2987
(1989).
\bibitem{BBOP} V.~Barger, M.~Berger, P.~Ohmann and R.~Phillips,
Phys. Rev. {\bf D50}, 4299 (1994).
\bibitem{ISAJET} F. Paige and S. Protopopescu, in {\it Supercollider Physics},
p. 41, ed. D. Soper (World Scientific, 1986);
H. Baer, F. Paige, S. Protopopescu and X. Tata,
in {\it Proceedings of the Workshop on Physics at Current Accelerators
and the Supercollider}, ed. J. Hewett, A. White and D. Zeppenfeld,
(Argonne National Laboratory pub. ANL-HEP-CP-93-92, 1993).
\bibitem{BKT} H.~Baer, C.~Kao and X.~Tata, Phys. Rev. {\bf D48}, R2978 (1993).
\bibitem{CTEQ} J. Botts {\it et. al.} Phys. Lett. {\bf B304}, 159 (1993).
\bibitem{BKTWINO} H. Baer, C.~Kao and X.~Tata, Phys. Rev. {\bf D48}, 5175
(1993).
\bibitem{KAMON}  T. Kamon, J.~Lopez, P.~McIntyre and
J.~T.~White, Texas A and M preprint, CTP-TAMU-19/94 (1994), have
argued that Drell-Yan + jet production, where a fluctuation causes
the jet to be misidentified as a lepton is the leading background to the
trilepton sample. We have not included this (detector-dependent) background
in Table 1, since it may be possible to eliminate it with a modest $\eslt$
cut with just a small loss of signal (except perhaps in the case where
$R$-parity conservation is broken by $B$-violating operators.
\bibitem{SUPERC} The impact of $R$-parity violation on the isolated
like-sign dilepton signal at hadron supercolliders has been discussed by
J. Gunion and P. Binetruy, Davis preprint, UCD-88-32 (1988) and by H.~Dreiner,
M.~Guchait and D.~P.~Roy, Phys. Rev. {\bf D49}, 3270 (1994).


\end{references}
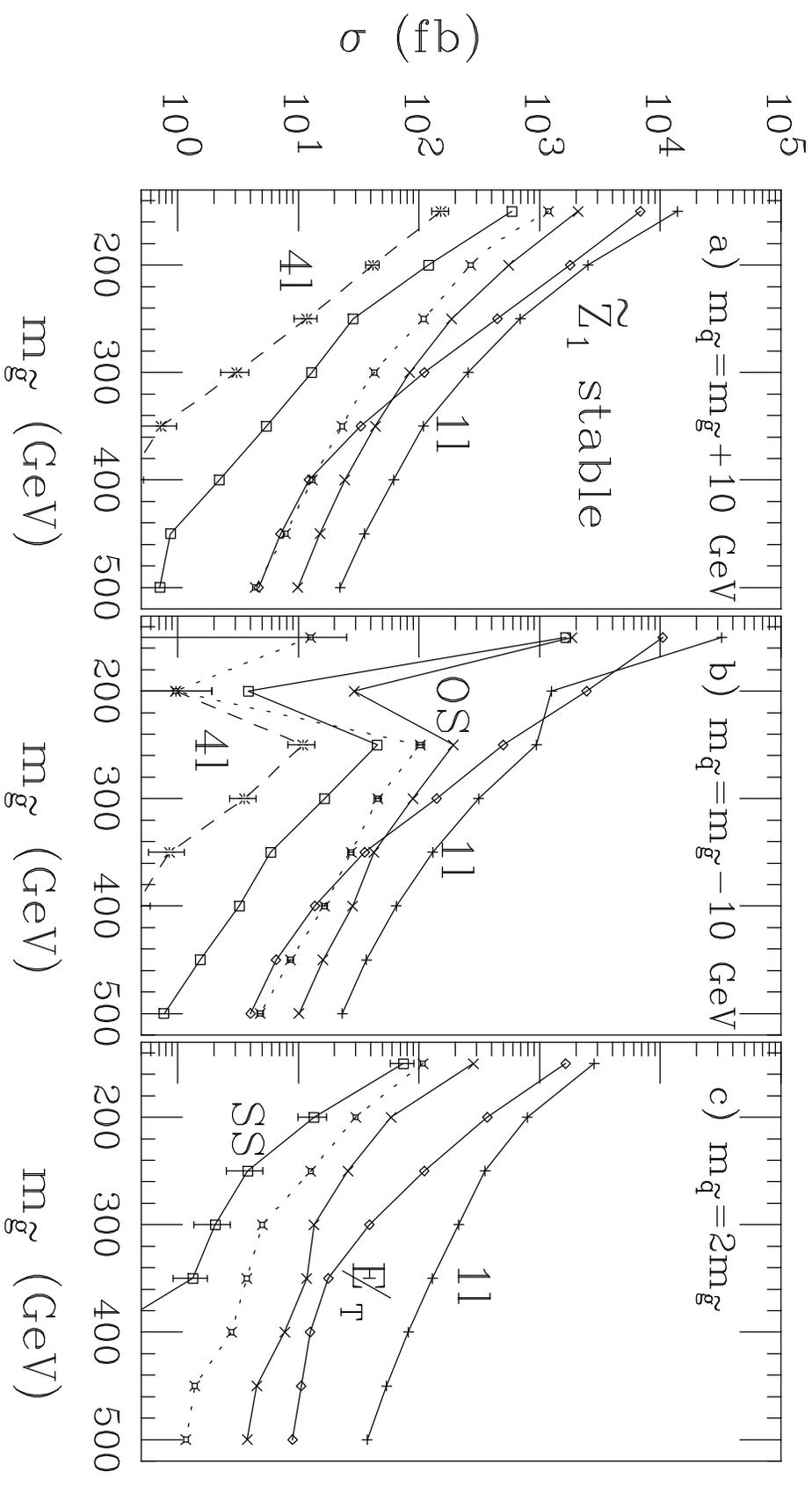
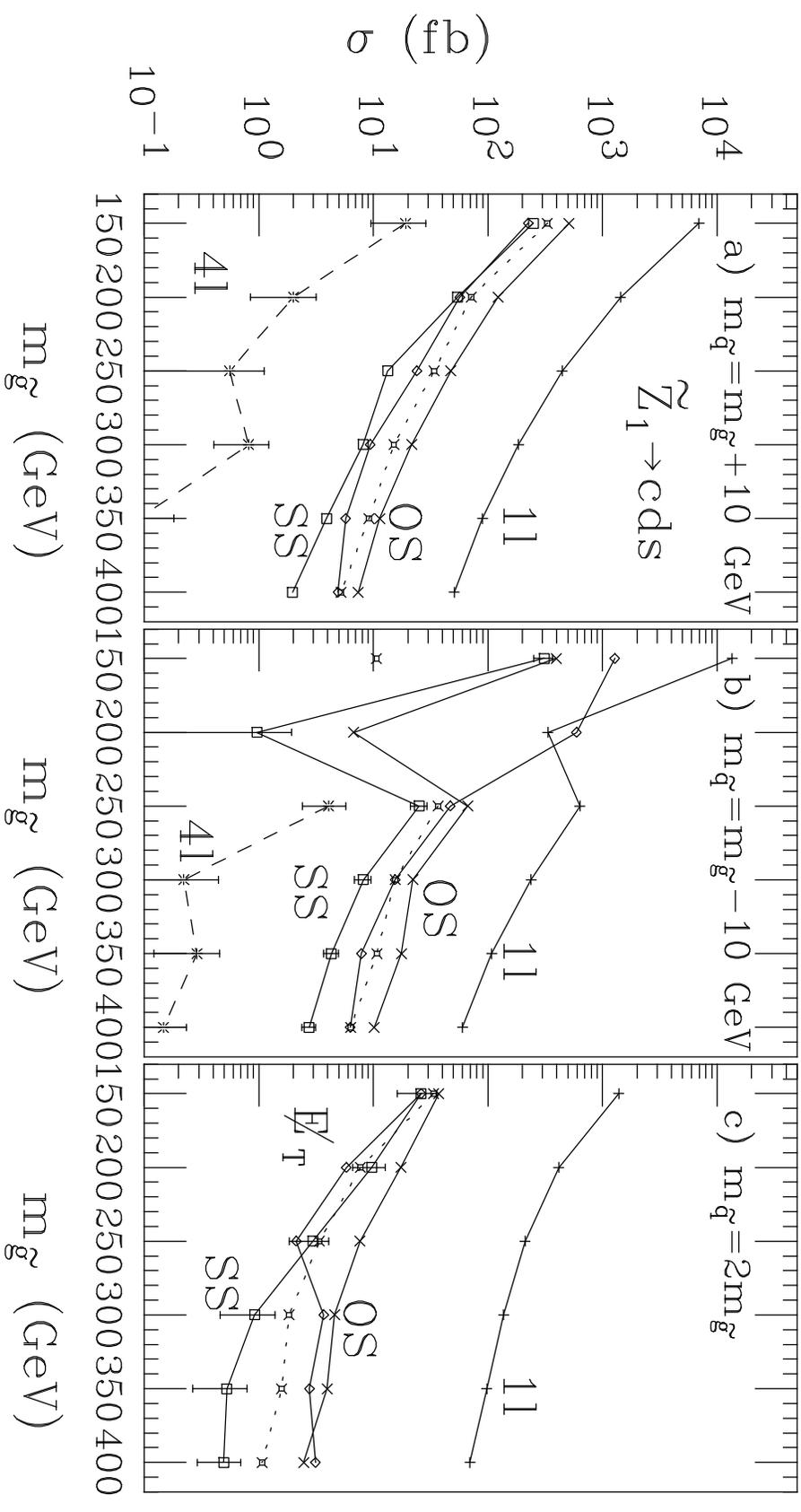
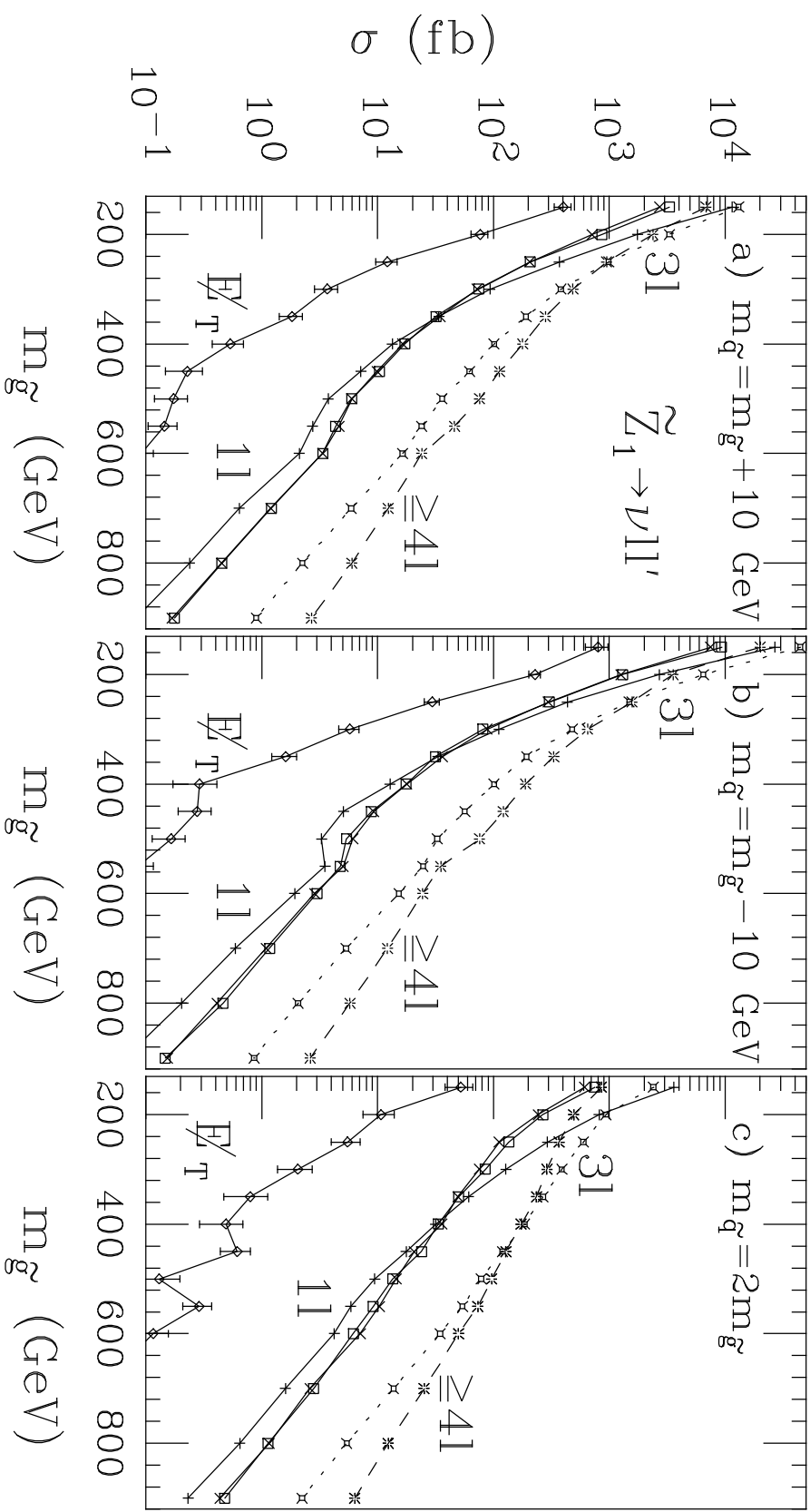
\end{document}